\documentclass[prb,twocolumn,showpacs,preprintnumbers,amsmath,aps]{revtex4}
\usepackage{graphicx,hyperref}
\usepackage{bm}
\usepackage{subfigure}

\newcommand\tr{\text{Tr}}

\def\nbC{{\mathchoice {\setbox0=\hbox{$\displaystyle\rm C$}%
\hbox{\hbox to0pt{\kern0.4\wd0\vrule height0.9\ht0\hss}\box0}}
{\setbox0=\hbox{$\textstyle\rm C$}\hbox{\hbox
to0pt{\kern0.4\wd0\vrule height0.9\ht0\hss}\box0}}
{\setbox0=\hbox{$\scriptstyle\rm C$}\hbox{\hbox
to0pt{\kern0.4\wd0\vrule height0.9\ht0\hss}\box0}}
{\setbox0=\hbox{$\scriptscriptstyle\rm C$}\hbox{\hbox
to0pt{\kern0.4\wd0\vrule height0.9\ht0\hss}\box0}}}}
\def\nbQ{{\mathchoice {\setbox0=\hbox{$\displaystyle\rm
Q$}\hbox{\raise
0.15\ht0\hbox to0pt{\kern0.4\wd0\vrule height0.8\ht0\hss}\box0}}
{\setbox0=\hbox{$\textstyle\rm Q$}\hbox{\raise
0.15\ht0\hbox to0pt{\kern0.4\wd0\vrule height0.8\ht0\hss}\box0}}
{\setbox0=\hbox{$\scriptstyle\rm Q$}\hbox{\raise
0.15\ht0\hbox to0pt{\kern0.4\wd0\vrule height0.7\ht0\hss}\box0}}
{\setbox0=\hbox{$\scriptscriptstyle\rm Q$}\hbox{\raise
0.15\ht0\hbox to0pt{\kern0.4\wd0\vrule height0.7\ht0\hss}\box0}}}}
\def\nbT{{\mathchoice {\setbox0=\hbox{$\displaystyle\rm
T$}\hbox{\hbox to0pt{\kern0.3\wd0\vrule height0.9\ht0\hss}\box0}}
{\setbox0=\hbox{$\textstyle\rm T$}\hbox{\hbox
to0pt{\kern0.3\wd0\vrule height0.9\ht0\hss}\box0}}
{\setbox0=\hbox{$\scriptstyle\rm T$}\hbox{\hbox
to0pt{\kern0.3\wd0\vrule height0.9\ht0\hss}\box0}}
{\setbox0=\hbox{$\scriptscriptstyle\rm T$}\hbox{\hbox
to0pt{\kern0.3\wd0\vrule height0.9\ht0\hss}\box0}}}}
\def\nbS{{\mathchoice
{\setbox0=\hbox{$\displaystyle     \rm S$}\hbox{\raise0.5\ht0%
\hbox to0pt{\kern0.35\wd0\vrule height0.45\ht0\hss}\hbox
to0pt{\kern0.55\wd0\vrule height0.5\ht0\hss}\box0}}
{\setbox0=\hbox{$\textstyle        \rm S$}\hbox{\raise0.5\ht0%
\hbox to0pt{\kern0.35\wd0\vrule height0.45\ht0\hss}\hbox
to0pt{\kern0.55\wd0\vrule height0.5\ht0\hss}\box0}}
{\setbox0=\hbox{$\scriptstyle      \rm S$}\hbox{\raise0.5\ht0%
\hboxto0pt{\kern0.35\wd0\vrule height0.45\ht0\hss}\raise0.05\ht0%
\hbox to0pt{\kern0.5\wd0\vrule height0.45\ht0\hss}\box0}}
{\setbox0=\hbox{$\scriptscriptstyle\rm S$}\hbox{\raise0.5\ht0%
\hboxto0pt{\kern0.4\wd0\vrule height0.45\ht0\hss}\raise0.05\ht0%
\hbox to0pt{\kern0.55\wd0\vrule height0.45\ht0\hss}\box0}}}}
\def\nbZ{{\mathchoice {\hbox{$\sf\textstyle Z\kern-0.4em Z$}}
{\hbox{$\sf\textstyle Z\kern-0.4em Z$}}
{\hbox{$\sf\scriptstyle Z\kern-0.3em Z$}}
{\hbox{$\sf\scriptscriptstyle Z\kern-0.2em Z$}}}}

\begin{document}

\title{Dimensional reduction and its breakdown in the 3-dimensional long-range random field Ising model}

\author{Maxime Baczyk} \email{baczyk@lptmc.jussieu.fr}
\affiliation{LPTMC, CNRS-UMR 7600, Universit\'e Pierre et Marie Curie,
bo\^ite 121, 4 Pl. Jussieu, 75252 Paris c\'edex 05, France}

\author{Matthieu Tissier} \email{tissier@lptmc.jussieu.fr}
\affiliation{LPTMC, CNRS-UMR 7600, Universit\'e Pierre et Marie Curie,
bo\^ite 121, 4 Pl. Jussieu, 75252 Paris c\'edex 05, France}

\author{Gilles Tarjus} \email{tarjus@lptmc.jussieu.fr}
\affiliation{LPTMC, CNRS-UMR 7600, Universit\'e Pierre et Marie Curie,
bo\^ite 121, 4 Pl. Jussieu, 75252 Paris c\'edex 05, France}

\author{Yoshinori Sakamoto} \email{yossi@phys.ge.cst.nihon-u.ac.jp}
\affiliation{Laboratory of Physics, College of Science and Technology, 
Nihon University, 7-24-1, Narashino-dai, Funabashi-city, Chiba, 
274-8501, Japan}

\date{\today}

\begin{abstract}
We investigate dimensional reduction, the property that the critical behavior of a system in the presence of quenched disorder in dimension $d$ 
is the same as that of its pure counterpart in $d-2$, and its breakdown in the case of the random-field Ising model in which both the interactions 
and the correlations of the disorder are long-ranged, {\it i.e.} power-law decaying. To some extent the power-law exponents play the role of 
spatial dimension in a short-range model, which allows us to probe the theoretically predicted existence of a nontrivial critical value separating 
a region where dimensional reduction holds from one where it is broken, while still considering the physical dimension $d=3$. By extending 
our recently developed approach based on a nonperturbative functional renormalization group combined with a supersymmetric formalism, 
we find that such a critical value indeed exists, provided one chooses a specific relation between the decay exponents of the interactions and 
of the disorder correlations. This transition from dimensional reduction to its breakdown should therefore be observable in simulations and 
numerical analyses, if not experimentally.

\end{abstract}


\maketitle

\section{Introduction}
\label{sec:introduction}

The random-field Ising model is a prototypical example of a system in which the presence of quenched disorder has a drastic effect on 
the collective behavior.\cite{imry-ma75,nattermann98} While long-range ferromagnetic order is still 
observed above a critical dimension, as in the 
pure Ising model, the properties of the associated critical point are strongly modified by the random field. In a renormalization 
group (RG) setting, temperature is irrelevant at the fixed point and the long-distance physics is therefore dominated 
by disorder-induced, sample-to-sample fluctuations rather than by thermal 
fluctuations.\cite{villain84,fisher86} This leads to anomalous scaling relations due to the existence of an additional critical exponent associated 
with temperature, to a shift of the lower and upper critical dimensions, a very slow spatial decay of the correlation functions at criticality, and a 
very strong slowing down of the dynamics close to the critical point that can be described in terms of an unconventional activated dynamic 
scaling.\cite{fisher86,fisher_activ,nattermann98}

One of the puzzles about the critical behavior of the RFIM was the property of \textit{dimensional reduction}, according to which the behavior in 
the presence of a random field is the same as that of the pure system in two dimensions less. This property, found to all orders in conventional 
perturbation theory\cite{aharony76,grinstein76,young77} and also nonperturbatively derived as a consequence of an 
underlying supersymmetry of the model at zero temperature\cite{parisi79,cardy83,klein83-84}, was proven to be wrong in 
$d=3$.\cite{imbrie84,bricmont87} From the supersymmetric 
approach it was understood that the failure was related to the presence of metastable states, {\it i.e.} 
of multiple extrema of the bare action (microscopic hamiltonian) in the region of interest.\cite{parisi84b} 
However, no further progress had been made.

We found a resolution of the dimensional-reduction puzzle by means of a \textit{nonperturbative functional RG} (NP-FRG) approach, showing 
that, as in the simpler case of a pinned interface in a random environment where a perturbative FRG 
analysis is sufficient,\cite{FRGfisher86,narayan-fisher92,balents96,chauve98,CUSPledoussal,CUSPledoussal09,CUSPledoussal11} 
breakdown of dimensional reduction is 
related to the appearance of a singularity in the functional dependence of the cumulants of the renormalized disorder, with however the 
singularity becoming too weak to cause a failure of dimensional reduction above a nontrivial 
critical dimension $d\simeq 5.1$.\cite{tarjus04,tissier06} 
More recently, we also showed that dimensional reduction breakdown is related to a spontaneous breaking of the underlying supersymmetry 
along the RG flow\cite{tissier11,tissier12,tissier12b} and that it is physically associated with the large-scale properties of the avalanches  
characterizing the behavior of the system at zero temperature.\cite{tarjus13}

Whereas the whole description obtained through the NP-FRG is consistent and leads to predictions, {\it e.g} for the critical exponents in $d=3$ 
and $d=4$, that are in good agreement with computer simulations and ground-state numerical studies, directly accessing the properties 
of the RFIM at and around the critical dimension of 5.1 is not feasible by computer studies (not to mention experiments!). The goal 
of the present work is to provide a way to get around this problem and to allow for a direct study of a $3$-dimensional system. To this end, 
we consider a RFIM with long-range interactions and long-range correlations of the random field. The interest in 
long-range models has a long history and comes from the fact that the presence 
of long-range, power-law decaying, interactions decreases the lower critical dimension of a model and that varying the exponent of the power 
law in a fixed dimension allows one to find a spectrum of critical behavior that goes  from mean-field for truly long-range interactions to the 
absence of transition for short-range decay while spanning a continuous range of nonclassical  behavior in between. In some sense, changing 
the exponent of the power law at fixed dimension is like changing the dimension in a short-range model.
 
To study dimensional reduction and its breakdown, one must introduce not only long-range interactions but also 
long-range correlations of the random field. As will be explained in more detail below, this is the only way to 
produce a supersymmetry in the field theory at zero temperature and therefore to possibly generate   
a dimensional reduction property. In addition, the exponents characterizing the decay of the interactions 
and of the disorder correlations have to be related in a specific manner. The problem can then be tackled through an 
extension of the NP-FRG approach combined with the 
supersymmetric formalism, which we have previously developed for the short-range RFIM.\cite{tissier11,tissier12,tissier12b} 
The main outcome of the theory is that there is a nontrivial critical value of the power exponent describing the spatial 
decay of the interactions that separate a domain 
where dimensional reduction is valid (for longer-range interactions) from a domain where it is not (for shorter-range interactions). 
This opens the way to a direct check of the transition between the presence and the absence of 
dimensional reduction in the critical behavior of the RFIM in the physically and technically accessible dimension $d=3$.

\section{Long-range model and supersymmetry}
\label{sec:SUSY_LR}

The model that we investigate is the field-theoretical version of the RFIM with long-range interactions and disorder correlations, 
with bare action (Hamiltonian)
\begin{equation}
\begin{split}
\label{eq_ham_dis}
&S[\varphi;h]=  S_B[\varphi]-\int_{x} h(x) \varphi(x) = \\&-\frac{1}{2}\int_{x}\int_{y}  \lambda(x-y)\varphi(x)\varphi(y)
+ \int_{x}\bigg\{ \frac{r}{2} \varphi(x)^2 + \frac{u}{4!} \varphi(x)^4  \\& -h(x) \varphi(x) \bigg\} ,
\end{split}
\end{equation}
where $ \int_{x} \equiv \int d^d x$ and the interaction goes as $\lambda(x-y)\sim \vert x-y\vert^{-(d+\sigma)}$ when $\vert x-y\vert \gg 1$, 
with $\sigma > 0$; $h(x)$ is a random source (a random magnetic field) that is taken with a Gaussian distribution characterized by a zero 
mean and a variance $\overline{h(x)h(y)}= \Delta(x-y) \sim \vert x-y\vert^{-(d-\rho)}$ when $\vert x-y\vert >> 1$. For $\sigma \geq 2$, 
one obviously recovers a model with short-range interactions and a similar reasoning applies  for $\rho\leq 0$.

Let us repeat the steps of the Parisi-Sourlas supersymmetric construction.\cite{parisi79} The critical behavior being controlled by a 
zero-temperature fixed point\cite{villain84,fisher86}, one can focus on the ground-state configuration which is solution 
of the stochastic field equation
\begin{equation}
\label{eq_stochastic_field_eq} 
\frac{\delta S[\varphi;h]}{\delta \varphi(x)}= J(x) \, ,
\end{equation}
where $J$ is an external source (a magnetic field) conjugate to the $\varphi$ field. When the solution is unique, 
which is precisely the crux of the problem and will be addressed later on, the 
equilibrium (Green's) correlation functions of the $\varphi$ field are obtained from the generating functional
\begin{equation}
  \label{eq_generating_func1}
\begin{aligned}
\mathcal Z_h[\hat{J},J]=\int& \mathcal D\varphi \; \delta\left[\dfrac{\delta S_B[\varphi]}{\delta \varphi}-h-J\right] 
\; \left|\det \dfrac{\delta^2 S_B[\varphi]}{\delta \varphi \delta \varphi}\right| \\& \times \exp  \int_{x}  \hat{J}(x)  \varphi(x) \,.
\end{aligned}
\end{equation}
Because of the assumed uniqueness of the solution, the absolute value of the determinant present in the right-hand side 
can be dropped and the functional can be built through standard field-theoretical techniques.\cite{zinnjustin89} 
One introduces  an auxiliary bosonic ``response'' field $\hat{\varphi}(x)$ to exponentiate the delta functional and 
two auxiliary fermionic ``ghost'' fields  $\psi(x)$ and  $\bar{\psi}(x)$ to exponentiate the determinant. In the resulting 
form, the average of the Gaussian random field can be explicitly performed and one obtains
\begin{equation}
\begin{aligned}
  \label{eq_generating_func3}
\mathcal Z&[\hat{J},J, \bar{K},K]= \overline{\mathcal Z_h[\hat{J},J, \bar{K},K]}\\&= 
\int \mathcal D\varphi \mathcal D\hat{\varphi} \mathcal D\psi D\bar{\psi} \exp \bigg\{-S_{ss}[\varphi,\hat{\varphi},\psi,\bar{\psi}] \, ,+ \\&  
\int_{x} \left( \hat{J}(x)  \varphi(x) + \psi(x)  \bar{K}(x)+K(x) \bar{\psi}(x) + J(x)  \hat{\varphi}(x)\right) \bigg\}
\end{aligned}
\end{equation}
where two fermionic sources, $\bar{K}(x),K(x)$, linearly coupled to the ghost fields have been introduced and
\begin{equation}
\label{eq_susyaction1}
\begin{aligned}
S_{ss}= & \int_{x}\hat{\varphi}(x) \dfrac{\delta S_B[\varphi]}{\delta \varphi(x)}   -  \int_{x}\int_{y}\bar{\psi}(x) 
\dfrac{\delta^2 S_B[\varphi]}{\delta \varphi(x) \delta \varphi(y)} \psi(y) \\&- \frac{1}{2}\int_{x}\int_{y}\hat{\varphi}(x)\Delta(x-y)\hat{\varphi}(y)\,.
\end{aligned}
\end{equation}
The $\varphi$-field connected correlation functions of the original problem are obtained from functional derivatives of 
$W[\hat{J},J, \bar{K},K]= \log\mathcal Z[\hat{J},J, \bar{K},K]$ with respect to $\hat{J}$ that are further evaluated for $K=\hat{K}=\hat{J}=0$.

The next step of the construction is to introduce a ``superspace'' by adding to the $d$-dimensional Euclidean space with coordinates 
$x=\left\lbrace x^\mu\right\rbrace $ two anti-commuting Grassmann coordinates $\theta,\bar{\theta}$ (satisfying 
$\theta^2=\bar{\theta}^2=\theta \bar{\theta}+\bar{\theta}\theta=0$)\cite{zinnjustin89}, so that the original field and the auxiliary 
fields can be grouped in a single ``superfield'' 
$\Phi(\underline{x})=\varphi(x) + \bar{\theta} \psi(x)+ \bar{\psi}(x) \theta + \bar{\theta}\theta \hat{\varphi}(x)$, 
where $\underline{x}=(x,\theta,\bar{\theta})$ denotes the coordinates in superspace. A similar procedure applies to 
the sources that can be grouped in a single ``supersource''  $\mathcal J(\underline x)$. At this stage, we leave 
unspecified the metric of the superspace. By using the 
properties of the Grassmann variables, Eq.~[\ref{eq_generating_func3}] can then be rewritten in the following compact form:
\begin{equation}
\begin{aligned}
  \label{eq_part_func_PS}
\mathcal Z[\mathcal J]=\int\mathcal D\Phi  \exp \left(-S_{ss}[\Phi] +   \int_{\underline x}  \mathcal J(\underline x)  \Phi(\underline x)\right) \, ,
\end{aligned}
\end{equation}
where
\begin{equation}
\begin{aligned}
\label{eq_susyaction2}
S_{ss}[\Phi]=& \int_{\underline x}\bigg \{\frac {r}{2}\Phi(\underline{x})^2+\frac {u}{4!}\Phi(\underline{x})^4\bigg \} + 
\frac 12 \int_{x}\int_{y} \int_{\underline \theta} \Phi(x,\underline{\theta}) \\&\times \big [-\lambda(x-y) - 
\Delta(x-y) \partial_{\theta}\partial_{\bar \theta}\big ]\Phi(y,\underline{\theta})
\end{aligned}
\end{equation}
and where $ \int_{\underline{\theta}} \equiv  \iint d\theta d\bar{\theta}$ and $ \int_{\underline{x}} \equiv \int_x  \int_{\underline{\theta}}$. 

To describe the long-distance physics, one needs information only about the low-momentum behavior of the Fourier transform of the 
long-range functions, $\tilde \lambda(q)$ and $\tilde \Delta(q)$, namely
\begin{equation}
\begin{aligned}
\label{eq_LRfunctionsFourier}
&\tilde \lambda(q)= \tilde \lambda(0) - Z_{LR} \, (q^2)^{\frac{\sigma}{2}} - Z\,  q^2 +\cdots \\&
\tilde \Delta(q)= \Delta_{LR} \, (q^2)^{-\frac{\rho}{2}}+\Delta + \cdots \, ,
\end{aligned}
\end{equation}
where the higher-order terms in $q^2$ indicated by the ellipses have been dropped as irrelevant. The above expressions have their 
counterpart in real space in terms of fractional derivatives.

For the short-range model, with $Z_{LR}=\Delta_{LR}=0$, terms in the interaction and in the disorder correlation can be combined to form a 
``super-Laplacian'' in superspace with an appropriately chosen metric.\cite{parisi79} To find the conditions under which this can be 
generalized to the long-range case, it proves more convenient to work in Fourier space for both the Euclidean and the Grassmann variables. 
After introducing $\eta, \bar{\eta}$ as the Grassmann ``momenta'' conjugate to the coordinates $\bar \theta,\theta$ (a standard definition of 
the Fourier transform in Grassmann space is used\cite{zinnjustin89}), we rewrite the action $S_{ss}[\Phi]$ in Eq.~(\ref{eq_susyaction2})  as
\begin{equation}
\begin{aligned}
\label{eq_susyaction3}
S_{ss}[\Phi]=&\int_{\underline x}U_B(\Phi(\underline{x})) + \frac 12 \int_{q}\int_{\underline \eta} \Phi(-q,\underline{\eta})
\bigg [Z_{LR} \, (q^2)^{\frac{\sigma}{2}} +\\& Z q^2 - \bar \eta \eta\left (\Delta_{LR} \, (q^2)^{-\frac{\rho}{2}}+\Delta \right )\bigg ]
\Phi(q,\underline{\eta})\, ,
\end{aligned}
\end{equation}
where $\int_q \equiv \int d^d q/(2\pi)^d$, $\underline \eta \equiv \{\eta, \bar{\eta}\}$, $\int_{\underline \eta} \equiv \iint d\eta d\bar{\eta}$, 
and $U_B(\Phi)=(\tau/2)\Phi^2 +(u/4!)\Phi^4$ with $\tau=r-\tilde \lambda(0)$.

Assume now an extension of the short-range case with a ``supermetric'' 
$d\underline{x}^2 = dx^{\mu} dx^{\mu} + C d\bar{\theta} d\theta $, with $C$ an unknown parameter to be determined, 
and the associated ``super-Laplacian'' $\Delta_{ss}= \partial_\mu \partial_\mu+(4/C) \partial_\theta \partial_{\bar{\theta}}$. The 
squared norm of a ``supermomentum'' $\underline q=\{q,\underline \eta \}$ is then given by $\underline q^2=q^2+(4/C)\eta \bar \eta$. 
It is now straightforward to check that the long-range components of the interaction, $Z_{LR} \, (q^2)^{\sigma/2}$, and of the 
disorder correlation, $\Delta_{LR} \, (q^2)^{-\rho/2}$, can be combined as a power of $\underline q^2$ if and only if 
\begin{equation}
\label{eq_rho_sigma}
\rho=2-\sigma
\end{equation}
and if the parameter $C$ of the supermetric is chosen appropriately. Then,
\begin{equation}
\begin{aligned}
\label{eq_susyLRmomentum}
Z_{LR} \, (\underline q^2)^{\frac{\sigma}{2}}&= Z_{LR} \, (q^2)^{\frac{\sigma}{2}}\left (1 + \eta \bar \eta\, \frac{4\sigma}{ 2 C}\, 
(q^2)^{-1}\right ) \\&
= Z_{LR} \, (q^2)^{\frac{\sigma}{2}} -\bar \eta \eta\, \frac{4  \sigma Z_{LR}}{ 2C}\, (q^2)^{-\frac{\rho}{2}},
\end{aligned}
\end{equation}
which corresponds to the long-range term in the action if one chooses $C=2 \sigma Z_{LR}/\Delta_{LR}$.

If one also considers the short-range contributions to the interaction and the disorder correlation, an additional condition is required, that 
relates $Z$ and $\Delta$ as
\begin{equation}
\label{eq_delta_Z}
\Delta= \left (\frac{2\, \Delta_{LR}}{ \sigma\, Z_{LR}}\right ) Z\,,
\end{equation}
and this can be generalized to include higher powers in momenta (see the conclusion). Under the above 
conditions, the action $S_{ss}[\Phi]$ can now be reexpressed as
\begin{equation}
\begin{aligned}
\label{eq_susyaction4}
&S_{ss}[\Phi]=\\&\int_{\underline x}U_B(\Phi(\underline{x})) + \frac 12 \int_{q}\int_{\underline \eta} \Phi(-q,\underline{\eta})
\bigg [Z_{LR} \, (\underline q^2)^{\frac{\sigma}{2}} + Z \underline q^2\bigg ]\Phi(q,\underline{\eta})
\end{aligned}
\end{equation}
which is the generalization to superspace and superfield of a $\varphi^4$ action in Euclidean space with long-range interaction and no disorder. 
(Note that when $Z_{LR}$ and $Z$ are different from zero, they can simply be set to $1$ by a simple rescaling of the fields 
and momenta.)

As in the short-range case, the above action is invariant under a large group of both bosonic and fermionic symmetries (the latter symmetries 
mixing bosonic and fermionic fields).\cite{tissier12} Of special importance is the supersymmetry associated with the orthosymplectic 
group OSp(2,d)\cite{osp2d} that contains the ``superrotations'' that preserve the metric of the superspace. As 
a result of the latter, it can be shown, both perturbatively\cite{parisi79} and nonperturbatively,\cite{cardy83,klein83-84} 
that the superfield theory with action $S_{ss}[\Phi]$ 
for a Euclidean dimension $d$ reduces to the simple field theory with action $S_{ss}[\varphi]$ in dimension $d-2$. if the superfield theory 
indeed correctly describes the critical behavior of the long-range RFIM this proves the dimensional-reduction property. One knows 
however that the Parisi-Sourlas construction breaks down when there are multiple solutions of the stochastic 
field equation.\cite{parisi84b,parisi82} This problem was previously resolved by two of us for the short-range RFIM\cite{tissier12,tissier12b} 
and we extend the proposed formalism to the long-range case below.

Before presenting the nonperturbative functional RG used to describe the long-distance physics, we recall a few known predictions of the 
critical behavior of the long-range RFIM. We need to first introduce a few definitions. As alluded to in the Introduction, the critical 
behavior of the RFIM, be it short or long range, is controlled by a zero-temperature fixed point. The renormalized temperature is therefore 
irrelevant and characterized by a critical exponent $\theta> 0$. As a result, the spatial decay of the correlations at criticality are described by 
two ``anomalous'' dimensions instead of one. The ``connected'' pair correlation (Green's) function behaves as\cite{nattermann98}
\begin{equation}
 \label{eq_conn_correl}
\overline{\langle\varphi(x) \varphi(y)\rangle-\langle\varphi(x)\rangle \langle\varphi(y)\rangle} \sim \frac{T}{\vert x-y\vert^{d-2+\eta}}
\end{equation}
whereas the ``disconnected'' one, which survives at zero temperature, behaves as
\begin{equation}
 \label{eq_disc_correl}
\overline{\langle\varphi(x)\rangle \langle\varphi(y)\rangle} \sim \frac{1}{\vert x-y\vert^{d-4+\bar\eta}}
\end{equation}
with $2\eta \geq \bar\eta \geq \eta$ and $\theta=2+\eta-\bar\eta$. 

Consider now the long-range RFIM in $d=3$ with $\rho=2-\sigma$, which represents the case of interest in the present study. From  
the results of Ref.~[\onlinecite{bray86}], one expects several regimes for the critical behavior of the model according to the value 
of the exponent $\sigma$:

(i) For $\sigma< 1/2$, a mean-field regime with classical exponents; $\sigma=1/2$ therefore plays the role of an upper critical dimension.

(ii) For $1/2 < \sigma <1$, a long-range regime with the anomalous dimensions fixed to $\bar\eta=\eta=2-\sigma$, but nontrivial values 
of the other critcal exponents. (This regime corresponds to the long-range exchange and random-field correlation regime in Bray's 
terminology.\cite{bray86})

(iii) For $\sigma>1$: no phase transition; $\sigma=1$ therefore plays the role of a lower critical dimension.

Contrary to the generic case studied by Bray, there are no other regimes, in particular no short-range regime, with the specific conditions 
$\rho=2-\sigma, d=3$. Dimensional reduction implies that the critical behavior of the model is the same as that of the pure, 
long-range interaction Ising model in $d=1$. For this latter system, one also expects that the ``upper critical'' value of $\sigma$ 
is $1/2$ and that the ``lower critical'' one is $1$, with no short-range regime. Note finally that the long-range RFIM in $d=2$ is not interesting for 
the present investigation, as, when the underlying superrotational invariance is satisfied, there is no range of $\sigma$ for which it displays 
a nonclassical (non mean-field) critical behavior.

\section{NP-FRG for the long-range model}
\label{sec:NPFRG_long-range}

The theoretical approach that we use for investigating the critical behavior of the long-range RFIM relies on the NP-FRG formalism 
previously developed for the short-range RFIM.\cite{tissier11,tissier12,tissier12b} It combines four main ingredients:

(1) \textit{A replica or multi-copy formalism}, in which the permutational symmetry among replicas is explicitly broken by introducing 
linear sources acting independently on each copy of the original system. Using expansions in the number of unconstrained (or ``free'') 
replica sums then gives us access to the cumulants of the renormalized disorder with their full functional dependence, which allows for
the emergence of a nonanalytic behavior in the field arguments.\cite{tarjus04,tissier06,tissier12}

(2) \textit{An extension of the parisi-Sourlas superfield construction in the presence of metastable states.} We introduce 
a weighting factor $\exp(-\beta S)$, involving an auxiliary temperature $\beta^{-1}$, to the solutions of the stochastic field 
equation Eq.~(\ref{eq_stochastic_field_eq}) when constructing the generating functional $\mathcal Z$. When $\beta^{-1}$ approaches 
$0$, only the ground state, \textit{i.e.} the state with minimum energy or bare action, contributes to the functional, as desired.

(3) \textit{An exact functional RG formalism}. It is a version of Wilson's continuous RG in which one follows the evolution of the 
``effective average action'',  which is  the generating functional of the $1$-particle irreducible (1PI) correlation functions. The flow with 
a running infrared (IR) scale $k$, from the bare action at the microscopic scale ($k=\Lambda$) to the full effective action at macroscopic 
scale ($k=0$), is governed by an exact RG equation.\cite{wetterich93,berges02}

(4) \textit{A nonperturbative supersymmetry-compatible approximation scheme for the effective average action}. It involves truncations in 
the ``derivative expansion'', \textit{i.e.} the expansion in the number of spatial derivatives of the fundamental fields, and in the ``expansion 
in number of free replica sums'',  \textit{i.e.} the cumulant expansion, and leads to a closed set of coupled NP-FRG equations that can be 
solved numerically.

Steps (1) and (2) above lead, via the usual field-theoretical techniques\cite{zinnjustin89} and an extension of the derivation in 
section~(\ref{sec:SUSY_LR}), to a superfield theory for an arbitrary number $n$ of copies in a curved superspace. The generating functional 
that generalizes Eq.~(\ref{eq_part_func_PS}) is then expressed as
\begin{equation}
\begin{aligned}
  \label{eq_part_func_multicopy}
&\mathcal Z^{(\beta)}[\{\mathcal J_a\}]\\&
= \int \prod_{a=1}^{n}\mathcal D\Phi_a  \exp \bigg(-S^{(\beta)}[\{\Phi_a\}] +  \sum_{a=1}^{n} \int_{\underline x} 
\mathcal J_a(\underline x)  \Phi_a(\underline x)\bigg),
\end{aligned}
\end{equation}
and the multicopy action is given by
\begin{equation}
\begin{aligned}
\label{eq_superaction_multicopy}
&S^{(\beta)}[\{\Phi_a\}] = \sum_{a} \int_{\underline{x}} 
\bigg \{ \frac 12 \Phi_a(\underline{x})\big [Z_{LR}\, (-\partial ^2)^{\sigma/2}-Z\, \partial ^2 \big ]\\& \Phi_a(\underline{x})
+U_{B}(\Phi_a(\underline{x}))\bigg \} -\sum_{a,b}\frac 12 \int_{x}
\int_{\underline{\theta}_1\underline{\theta}_2}\bigg \{ \Phi_{a}(x,\underline{\theta}_1)\\&
\big [\Delta_{LR}\, (-\partial ^2)^{-1+\sigma/2}-\Delta\, \partial ^2\big]\Phi_{b}(x,\underline{\theta}_2)
+ perm(12)\bigg \} \, ,
\end{aligned}
\end{equation}
where $(-\partial ^2)^{\alpha}$ with $\alpha$ a real number is a symbolic notation describing a fractional Laplacian in Euclidean 
space; its Fourier transform generates a term in $(q^2)^{\alpha}$ (and for $\alpha=1$ one recovers the standard Laplacian). In the 
above equations, we have introduced a superspace whose $2$-dimensional Grassmannian subspace is curved, with the curvature 
proportional to $\beta$. For instance, the integral over Grassmannian subspace is defined as $\int_{\underline \theta}\equiv \int 
\int (1+\beta \bar\theta \theta)d\theta d\bar\theta$.\cite{tissier12} As discussed in detail in Ref.~[\onlinecite{tissier12}], 
the action in Eq. (\ref{eq_superaction_multicopy}) is still invariant under a large group of symmetries and supersymmetries.

We have applied the NP-FRG formalism to this superfield theory [step (3) above]. This proceeds by first 
introducing an infrared (IR) regulator that enforces a progressive account of the fluctuations to the bare action,
\begin{equation}
\begin{aligned}
\label{eq_regulator}
\Delta S_k^{(\beta)}=&\frac 12  \sum_{a}\int_{x_1 x_2}\int_{\underline \theta}
\Phi_{a}(x_1,\underline \theta)\widehat{R}_k(|x_1-x_2|)\Phi_{a}(x_2,\underline{\theta})\\&
+\frac 12 \sum_{a,b}\int_{\underline{x}_1 \underline{x}_2} \Phi_{a}(\underline{x}_1)
\widetilde{R}_k(|x_1-x_2|) \Phi_{b}(\underline{x}_2),
\end{aligned}
\end{equation}
with two IR cutoff functions $\widehat{R}_k$ and $\widetilde{R}_k$ suppressing the integration over modes with momentum 
$\vert q \vert \ll k$ (but not for those with $\vert q \vert \gg k$) \cite{berges02,tarjus04} in the modified $k$-dependent 
generating functional
\begin{equation}
\begin{aligned}
  \label{eq_part_func_multicopy}
\mathcal Z_k^{(\beta)}[\{\mathcal J_a\}]
= &\int \prod_{a=1}^{n}\mathcal D\Phi_a  \exp \bigg(-S^{(\beta)}[\{\Phi_a\}] + \\& \sum_{a=1}^{n} \int_{\underline x} 
\mathcal J_a(\underline x)  \Phi_a(\underline x)  -\Delta S_k^{(\beta)}[\{\Phi_a\}] \bigg) \, .
\end{aligned}
\end{equation}
We have chosen the two IR cutoff functions to be related through
\begin{equation}
\label{eq_cutoffs_relation}
\widetilde{R}_k(q^2)=-\left (\frac{2\, \Delta_{LR}}{ \sigma\, Z_{LR}}\right )\partial_{q^2}\widehat{R}_k(q^2)\, .
\end{equation}
The above relation and the form of the regulator ensure that all symmetries and supersymmetries of the theory  
are satisfied. This includes the superrotational invariance found (only) when the theory is restricted to a \textit{single} copy 
and to an infinite auxiliary temperature, $\beta=0$.\cite{tissier12} A specific form for the cutoff function $\widehat{R}_k$ 
will be given below.

We next introduce the effective average action,\cite{wetterich93,berges02} which is obtained from $\log \mathcal Z_k^{(\beta)}$ 
through a (modified) Legendre transform: 
\begin{equation}
\begin{split}
\label{eq_effective_average_action}
&\Gamma_k^{(\beta)}[\{\Phi_a\}]=\\& -\log \mathcal Z_k^{(\beta)}
[\{\mathcal J_a\}]+\sum_a\int_{\underline x} \Phi_a(\underline x) \mathcal J_a(\underline x) -\Delta S_k^{(\beta)}[\{\Phi_a\}]\,.
\end{split}
\end{equation}
As already mentioned, it is the generating functional of the 1PI correlation functions,\cite{zinnjustin89}  and it is the central 
quantity of our NP-FRG approach. Its dependence on the IR cutoff $k$ is governed by an exact renormalization-group 
equation (ERGE),\cite{berges02}
\begin{equation}
\label{eq_ERGE}
\partial_t \Gamma_k^{(\beta)}[\left\lbrace \Phi_a \right\rbrace]=\frac 12 \tr  \left\lbrace \partial_t \mathcal R_{k} \;\mathcal P_{k}^{(\beta)}
[\left\lbrace \Phi_a \right\rbrace]\right\rbrace ,
\end{equation}
where $t=\log(k/\Lambda)$ and the trace involves summing over copy indices and integrating over superspace; the modified 
propagator $\mathcal P_{k,ab}^{(\beta)}(\underline{x}_1,\underline{x}_2)$ is the (operator) inverse of 
$([\Gamma_k^{(\beta)}]^{(2)} +  \mathcal R_k)$ where $[\Gamma_k^{(\beta)}]^{(2)}[\left\lbrace \Phi_a \right\rbrace] $ is the 
second functional derivative of the effective average action with respect to the superfields $\Phi_a(\underline{x})$. 

The effective average action can be expanded in increasing number of unrestricted sums over copies, which generates an 
analog of a cumulant expansion for the renormalized disorder (more details are found in Refs.~[\onlinecite{tarjus04,tissier12}]): 
\begin{equation}
\label{eq_free_replica_sums}
\Gamma_k^{(\beta)}\left[\{ \Phi_a\}\right ]=
\sum_a \Gamma_{k1}^{(\beta)}[\Phi_a]-\frac{1}{2}\sum_{a,b}\Gamma_{k2}^{(\beta)}[\Phi_a,\Phi_b] + \cdots
\end{equation}
where (with a pinch of salt, see Refs.~[\onlinecite{tarjus04,tissier12}]), $\Gamma_{kp}^{(\beta)}$ the $p$th cumulant of the 
renormalized disorder at the scale $k$. Such expansions in increasing number of free sums over copies lead to systematic 
algebraic manipulations that allow one to derive a hierarchy of coupled ERGE's for the cumulants of the renormalized disorder 
from the ERGE for $\Gamma_k^{(\beta)}[\{\Phi_a\}]$, Eq.~(\ref{eq_ERGE}).

In Refs.~[\onlinecite{tissier12,tissier12b}], we showed that the ground state dominance when $\beta \rightarrow \infty$ comes 
with a formal property of the random generating functional, which was termed ``Grassmannian ultralocality". This property is also 
asymptotically found for finite $\beta$ when $k\rightarrow 0$ (after going to dimensionless quantities, \textit{i.e.} at the fixed point). 
When it is satisfied, the ERGE's for the cumulants simplify. They only involve ''ultralocal" parts of the cumulants, \textit{e.g.},
\begin{equation}
\label{eq_flow_Gamma1_ULapp}
\begin{split}
&\Gamma_{k1}^{(\beta)}[\Phi]=\int_{\underline \theta} \Gamma_{k1}[\Phi(\underline \theta)]\\&
\Gamma_{k2}^{(\beta)}[\Phi_1,\Phi_2]=\int_{\underline \theta_1}\int_{\underline \theta_2} \Gamma_{k2}
[\Phi_1(\underline \theta_1),\Phi_2(\underline \theta_2)] \, ,
\end{split}
\end{equation}
etc, where, in the right-hand sides, $\Gamma_{k1}$, $\Gamma_{k2}$, $\cdots$, only depend on the 
superfields at the explicitly displayed ``local" Grassmannian  coordinates (on the other hand, the dependence 
on the Euclidean coordinates, which is left implicit, still involves derivatives). $\Gamma_{k1}$, $\Gamma_{k2}$, $\cdots$, are 
then shown to be independent of the auxiliary temperature $\beta^{-1}$,
and the corresponding ERGE's can be further evaluated for physical fields $\Phi_a(\underline x)=\phi_a(x)$, \textit{i.e.} for 
superfields that are uniform in the Grassmann subspace.\cite{tissier12} For instance, one obtains 
\begin{equation}
\label{eq_flow_Gamma1_ULapp}
\begin{split}
&\partial_t\Gamma_{k1}\left[\phi_1\right ]= \\&-
\dfrac{1}{2} \tilde{\partial}_t \int_{x_2x_3}\widehat{P}_{k;x_2x_3}[\phi_1] \big(\Gamma_{k2;x_2,x_3}^{(11)}\left[\phi_1,\phi_1\right ] 
- \widetilde{R}_{k;x_2x_3}\big)
\end{split}
\end{equation}
and
\begin{equation}
\label{eq_flow_Gamma2_ULapp}
\begin{split}
&\partial_t\Gamma_{k2}\left[\phi_1,\phi_2\right ]=\\&
\dfrac{1}{2} \tilde{\partial}_t \int_{x_3x_4}\big \{- \widehat{P}_{k;x_3x_4}\left[\phi_1\right ] \Gamma_{k3;x_3,.,x_4}^{(101)}
\left[\phi_1,\phi_2,\phi_1\right ]+\\& \widetilde{P}_{k;x_3x_4}\left[\phi_1,\phi_1\right ] \Gamma_{k2;x_3x_4,.}^{(20)}
\left[\phi_1,\phi_2\right ]+ \frac{1}{2}\widetilde{P}_{k;x_3x_4}\left[\phi_1,\phi_2\right ] \\& \times \left( \Gamma_{k2;x_3,x_4}^{(11)}
\left[\phi_1,\phi_2\right ] - \widetilde{R}_{k;x_3x_4}\right) + perm(12)\big \},
\end{split}
\end{equation}
where $perm (12)$ denotes the expression obtained by permuting $\phi_1$ and $\phi_2$,  $\widetilde{\partial}_t$ 
is a short-hand notation to indicate a derivative acting only on the cutoff functions (\textit{i.e.},  $\widetilde{\partial}_k \equiv 
\partial_k \widehat{R}_k\, \delta/\delta \widehat{R}_k + \partial_k \widetilde{R}_k \, \delta/\delta \widetilde{R}_k$), and 
superscripts indicate functional differentiation with respect to the field arguments. The 
propagators $\widehat{P}_{k}$ and $\widetilde{P}_{k}$ are defined as
\begin{equation}
\label{eq_hatP_zero}
\widehat {P}_{k}[\phi ]=\left(\Gamma _{k,1}^{(2)}[ \phi ]+\widehat R_k\right) ^{-1}
\end{equation}
and
\begin{equation}
\label{eq_tildeP_zero}
\widetilde {P}_{k}[\phi_1, \phi_2 ]= \widehat {P}_{k}[ \phi_1 ](\Gamma _{k,2}^{(11)}[\phi_1, \phi_2 ]
-\widetilde R_k ) \widehat {P}_{k}[ \phi_2 ] \,.
\end{equation}

Finally, we use the fact that the continuous symmetries and supersymmetries of the theory lead to 
Ward-Takahashi identities\cite{tissier11,tissier12} at each running scale $k$. Important ones are 
those associated with superrotational invariance when the multi-copy theory is restricted to a one-copy theory by an 
appropriate choice of the sources and the auxiliary temperature has dropped from the formalism.\cite{tissier12} 
In particular, one then has for uniform field configurations,
\begin{equation}
\label{eq_ward-takahashi}
\Gamma _{k2}^{(11)}(q^2;\phi, \phi)=\left (\frac{2\, \Delta_{LR}}{\sigma\, Z_{LR}}\right )
\partial_{q^2}\Gamma _{k2}^{(2)}(q^2;\phi) \, . 
\end{equation}
Note that the relation in Eq.~(\ref{eq_cutoffs_relation}) between the two IR cutoff functions is precisely of the same form. 
So long as superrotational invariance is not spontaneously broken and Eq.~(\ref{eq_ward-takahashi}) remains valid, the ERGE 
for the first cumulant $\Gamma_{k1}$ can be closed thanks to the above Ward-Takahashi identities 
[see Eq.~(\ref{eq_flow_Gamma1_ULapp})]. It exactly reduces to that for the effective average action in the pure model 
in dimension $(d-2)$, which entails the dimensional-reduction property.\cite{tissier11,tissier12}

\section{Supersymmetry-compatible approximation scheme}
\label{sec:SUSYcompatible}

To actually solve the ERGE's describing the critical behavior of the long-range RFIM [see Eqs.~(\ref{eq_flow_Gamma1_ULapp}) 
and (\ref{eq_flow_Gamma2_ULapp})], we use the supersymmetry-compatible nonperturbative approximation scheme that we have 
already introduced in previous work.\cite{tissier11,tissier12} It combines a truncation in the derivative expansion, which approximate 
the long-distance behavior of the 1PI vertices, and a truncation in the expansion in cumulants of the renormalized disorder, while ensuring that 
the Ward-Takahashi identities associated with the supersymmetry are not explicitly violated. In the present case, the 
derivative expansion must be generalized to account for the long-range spatial decay of the interactions and the disorder correlations.
The minimal truncation then reads
\begin{equation}
\begin{aligned}
\label{eq_ansatz_gamma1}
&\Gamma_{k1}[\phi]=\\& \int_{x}\bigg \{U_k(\phi(x))+\frac{1}{2}Z_{LR,k}(\phi(x))\phi(x)(-\partial^2)^{\sigma/2}\phi(x)\\&
+ \frac{1}{2}Z_k(\phi(x))(\partial_{\mu}\phi(x))^2 \bigg \},
\end{aligned}
\end{equation}
\begin{equation}
\begin{aligned}
\label{eq_ansatz_gamma2}
&\Gamma_{k2}[\phi_1,\phi_2]=\\&\int_{x}\bigg \{\frac 12 \Delta_{LR,k}(\phi_1(x),\phi_2(x))
\big [\phi_1(x)(-\partial^2)^{-1+\sigma/2}\phi_2(x)
\\& +\phi_2(x)(-\partial^2)^{-1+\sigma/2}\phi_1(x) \big ] + V_k(\phi_1(x),\phi_2(x))\bigg \},
\end{aligned}
\end{equation}
with the higher-order cumulants set to zero.

Inserted in the ERGE's for the cumulants, Eqs.~(\ref{eq_flow_Gamma1_ULapp}) and (\ref{eq_flow_Gamma2_ULapp}), 
the above ansatz provides $5$ coupled flow equations for the $1$-copy potential $U_k(\phi)$ that describes the thermodynamics 
of the system, the two field-renormalization functions $Z_k(\phi)$ and $Z_{LR,k}(\phi)$, the $2$-copy potential $V_k(\phi_1,\phi_2)$ 
[from which one obtains the local part of second cumulant of the renormalized random field 
$\Delta_k(\phi_1,\phi_2) =\partial _{\phi_1}\partial _{\phi_2}V_k(\phi_1,\phi_2)$] and the strength of the long-range component of 
the second cumulant of the random field $\Delta_{LR,k}(\phi_1,\phi_2)$.

The NP-FRG flow equations are supplemented by an initial condition at the microscopic (UV) scale $k=\Lambda$. It is given by the bare 
action that can be recast as
\begin{equation}
\begin{aligned}
\label{eq_ansatz_gamma1UV}
\Gamma_{\Lambda1}[\phi]= \int_{x}\bigg \{&U_B(\phi(x))+\frac{1}{2}\phi(x)(-\partial^2)^{\sigma/2}\phi(x)\\&
+ \frac{Z}{2}(\partial_{\mu}\phi(x))^2 \bigg \},
\end{aligned}
\end{equation}
where, without loss of generality, we have set $Z_{LR}=1$, and
\begin{equation}
\begin{aligned}
\label{eq_ansatz_gamma2UV}
\Gamma_{\Lambda2}[\phi_1,\phi_2]&=\int_{x}\bigg \{\frac {\Delta_{LR}}{2}
\big [\phi_1(x)(-\partial^2)^{-1+\sigma/2}\phi_2(x)
\\& +\phi_2(x)(-\partial^2)^{-1+\sigma/2}\phi_1(x) \big ] + \Delta \phi_1(x)\phi_2(x)\bigg \}
\end{aligned}
\end{equation}
with $\Delta_{LR}>0$. To ensure that supersymmetry is not violated by the initial action, we choose 
$\Delta=(2/\sigma)Z \Delta_{LR}$ [see Eq.~(\ref{eq_delta_Z})].

The first observation is that due to the long-range nature of the interaction and disorder correlation, characterized by a 
nontrivial exponent $\sigma$, the propagators $\widehat {P}_{k}[\phi ]$ and  $\widetilde {P}_{k}[\phi_1, \phi_2 ] $ 
appearing in the flow equations have a nonanalytic momentum dependence, even away from criticality. By introducing 
the above ansatz, Eqs.~(\ref{eq_ansatz_gamma1},\ref{eq_ansatz_gamma2}), in Eqs.~(\ref{eq_hatP_zero}) and 
(\ref{eq_tildeP_zero}), one finds for uniform field configurations,
\begin{equation}
\begin{aligned}
\label{eq_hatP_ansatz}
(\widehat {P}_{k}[\phi])_{qq'}&=\delta(q+q')\widehat {P}_{k}(q^2;\phi)
\end{aligned}
\end{equation}
and
\begin{equation}
\begin{aligned}
\label{eq_tildeP_ansatz}
&(\widetilde {P}_{k}[\phi_1, \phi_2 ])_{q q'}=\delta(q+q')\bigg ((q^2)^{-1+\frac{\sigma}{2}}\big [\Delta_{LR,k}(\phi_1, \phi_2) +\\&
\frac 12 (\phi_1\partial\phi_1+\phi_2\partial\phi_2)
\Delta_{LR,k}(\phi_1, \phi_2)\big ]+ \Delta_k(\phi_1, \phi_2)-\widetilde R_k(q^2) \bigg )\\& \widehat {P}_{k}(q^2;\phi_1)
\widehat {P}_{k}(q^2;\phi_2)\, ,
\end{aligned}
\end{equation}
with
\begin{equation}
\begin{aligned}
\label{eq_hatP_ansatz_unif}
&\widehat {P}_{k}(q^2;\phi)=\\& \frac{1}{\partial_{\phi}[\phi Z_{LR,k}(\phi)](q^2)^{\sigma/2} + Z_k(\phi)q^2 + U''_k(\phi)
+\widehat R_k(q^2)}\,.
\end{aligned}
\end{equation}
The RG flow of $Z_{LR,k}$ and $\Delta_{LR,k}$ is obtained by extracting from the ERGE of the relevant two-point 1PI vertex, 
$\Gamma_{k,1}^{(2)}$ for the former and $\Gamma_{k,2}^{(11)}$ for the latter, that part which has the proper singular 
momentum dependence (when evaluated for uniform fields). One finds as a result that the flow of the two functions $Z_{LR,k}$ and 
$\Delta_{LR,k}$ involves field derivatives of themselves in such a way that if $Z_{LR,k}$ and $\Delta_{LR,k}$ are independent of the fields 
at the UV scale, which is indeed the case here [see Eqs~(\ref{eq_ansatz_gamma1UV},\ref{eq_ansatz_gamma2UV})], $Z_{LR,k}$ and 
$\Delta_{LR,k}$ do not flow and remain equal to their bare values. This is in line with the conclusions of Ref.~[\onlinecite{bray86}].

In the present $3$-dimensional case, any crossover to short-range behavior could only occur for $\sigma>1$: indeed, the latter behavior is 
predicted for $\sigma>2-\eta_{SR}$ and/or $\sigma>2-2\eta_{SR}+\bar \eta_{SR}$,\cite{bray86} which, in the present random-field system 
where $\eta_{SR} \approx 2\eta_{SR}-\bar \eta_{SR}\approx 0.5$ in $d=3$,\cite{nattermann98,tissier12} only takes place in the 
region where no phase transition is observed anyhow. The whole $\sigma$ range of interest, \textit{ i.e.} $1/2<\sigma<1$ 
(see the Introduction), is therefore in the long-range regime characterized by fixed anomalous dimensions given by 
$\eta=\bar\eta=2-\sigma$. The latter result can be easily understood by considering the propagators in 
Eqs.~(\ref{eq_hatP_ansatz},\ref{eq_tildeP_ansatz},\ref{eq_hatP_ansatz_unif}). When $k\rightarrow 0$ at the critical point, 
$\widehat {P}_{k\rightarrow 0}(q^2;\phi)$ reduces to the connected pair correlation function in Eq.~(\ref{eq_conn_correl}) while 
$\widetilde {P}_{k\rightarrow 0}(q^2;\phi,\phi)$ reduces to the disconnected pair correlation function in Eq.~(\ref{eq_disc_correl}). 
The low-momentum behavior in Eqs.~(\ref{eq_hatP_ansatz},\ref{eq_tildeP_ansatz},\ref{eq_hatP_ansatz_unif}) is dominated 
by the singular terms, which are both in $(q^2)^{-\sigma/2}$. Comparison with the definitions of the anomalous dimensions 
in Eqs.~(\ref{eq_conn_correl},\ref{eq_disc_correl}) then directly provides the result. 

To cast the NP-FRG flow equations in a dimensionless form that allows one to describe the long-distance physics associated with 
the critical point, one must introduce scaling dimensions that account for the fact that the fixed point is at  ``zero-temperature''. 
Near such a  fixed point, the renormalized temperature is irrelevant and characterized by an exponent $\theta>0$, and one has the 
following scaling dimensions (see also section~\ref{sec:SUSY_LR}):
\begin{equation}
\begin{aligned}
\label{eq_scaling_dimension}
&Z_{k} \sim k^{-\eta}, \; \phi_a  \sim k^{\frac{1}{2}(d-4+\bar \eta)},\\&
U_k\sim k^{d-\theta}, \;V_k \sim k^{d-2\theta},
\end{aligned}
\end{equation}
so that the local component of the second cumulant of the renormalized random field $\Delta_k$ scales as 
$k^{-(2\eta- \bar \eta)}$. As discussed above,  $\eta=\bar\eta=2-\sigma$, so that $\theta=2$,  in the long-range regime considered 
here (with $1/2<\sigma<1$ in $d=3$). Note that, in contrast with the short-range RFIM, the equality of $\eta$ and $\bar\eta$ 
and a fixed value of the temperature exponent $\theta=2$ do \textit{not} necessarily mean that dimensional reduction is obeyed.

Recalling that the long-range functions are \textit{not} renormalized and using lower-case letters, 
$u_k, v_k,\delta _k,  \varphi$, to denote the dimensionless counterparts of 
$U_k, V_k,\Delta _k,  \phi$,  the dimensionless form of the flow equations can be symbolically written as
\begin{equation}
\label{eq_flow_dimensionless}
\begin{split}
&\partial_t u'_k(\varphi)=\beta_{u'}(\varphi),\\&
\partial_t z_k(\varphi)=\beta_{z}(\varphi),\\&
\partial_t \delta_k(\varphi_1,\varphi_2)=\beta_{\delta}(\varphi_1,\varphi_2),
\end{split}
\end{equation}
where a prime denotes a derivative with respect to the argument. The beta functions themselves 
depend on $u_k'$, $z_k$, $\delta_k$ and their derivatives. They are obtained from the expressions for the short-range RFIM 
given in Ref.~[\onlinecite{tissier12b}] after the following replacements:
\begin{equation}
\begin{aligned}
\label{eq_replacementLR_SR}
&z_k(\varphi)y + u''_k(\varphi)+s(y) \rightarrow y^{\sigma/2} + z_k(\varphi)y+ u''_k(\varphi)+s(y),\\&
 \delta_k(\varphi_1,\varphi_2)+s'(y) \rightarrow  y^{-1+\sigma/2} + \delta_k(\varphi_1,\varphi_2)+s'(y),
\end{aligned}
\end{equation}
where $y=q^2/k^2$ is the rescaled squared momentum and $s(y)$ is the dimensionless form of the IR cutoff function that is introduced 
through
\begin{equation}
\begin{aligned}
\label{eq_cutoffdimensionless}
&\widehat R_k(q^2)=k^2 s(q^2/k^2) \\&
\widetilde R_k(q^2) = - \frac{2}{\sigma}\Delta_{LR} \partial_{q^2}\widehat R_k(q^2)= -\frac{2}{\sigma}\Delta_{LR}\,  s'(q^2/k^2)\, .
\end{aligned}
\end{equation}
Note that even if one chooses an initial condition where the short-range components of the interaction and the disorder 
correlation are zero, $Z=\Delta=0$ [which still satisfies Eq.~(\ref{eq_delta_Z})], these components are generated along the RG flow.

As the chosen ansatz for the renormalized cumulants and for the IR cutoff functions do not \textit{explicitly} violate the 
Ward-Takahashi identities associated with superrotational invariance, we find the same property as in the short-range 
RFIM:\cite{tissier12} so long as the local piece of the cumulant of the renormalized random field $\delta_k(\varphi_1,\varphi_2)$ 
is well enough behaved when $\varphi_2 \rightarrow \varphi_1$, \textit{i.e.} does not develop a cusp in $\vert \varphi_2 - \varphi_1\vert$, 
superrotational invariance is not \textit{spontaneously} broken along the RG flow. One then exactly finds 
that $\delta_k(\varphi,\varphi)=z_k(\varphi)$ and that the flows of $z_k(\varphi)$ and $u'_k(\varphi)$ are 
identical to those obtained in the NP-FRG treatment of the pure model with long-range interactions in dimension $d-2$. 
Two situations may be encountered. Either this remains true down to the $k \rightarrow 0$ limit, and the fixed point 
describes a critical behavior with dimensional reduction, or a cusp appears at a specific (``Larkin'') scale $k_L>0$ along 
the flow and dimensional reduction must be broken.

\section{Results in the three-dimensional case and discussion}
\label{sec:results}

The main goal of this work is to investigate in the $3$-dimensional long-range RFIM the existence of a critical value $\sigma_{DR}$ 
separating a region of parameter $\sigma$ where dimensional reduction holds from a region where it does not. We have 
thus looked for the signature of the appearance of a cusp along the RG flow, which corresponds to the disappearance of the 
dimensional-reduction fixed point. 
To this end, we have studied the second derivative of $\delta_k(\varphi_1,\varphi_2)$ with respect to $\varphi_1$ and $\varphi_2$ 
when evaluated in the limit $\varphi_2 \rightarrow \varphi_1$:
\begin{equation}
\label{eq_deltak2}
 \delta_{k,2}(\varphi)=-\partial_{\varphi_1}\partial_{\varphi_2} \delta_k(\varphi_1,\varphi_2)\bigg \vert_{\varphi_1=\varphi_2=\varphi}.
\end{equation}
In the absence of a cusp,  $\delta_{k,2}(\varphi)$ is finite whereas it blows up whan a cusp first appears.

The flow of $\delta_{k,2}(\varphi)$ is simply obtained from the one for $\delta_k(\varphi_1,\varphi_2)$ 
[see Eq.~(\ref{eq_flow_dimensionless})] by using Eq.~(\ref{eq_deltak2}) and assuming that
\begin{equation}
\label{eq_expansion_delta}
\delta_k(\varphi_1,\varphi_2)=\delta_{k,0}(\varphi)+(1/2)\delta_{k,2}(\varphi)(\varphi_1-\varphi_2)^2 + \cdots\, ,
\end{equation}
with $\varphi=(1/2)(\varphi_1+\varphi_2)$, when $\varphi_2 \rightarrow \varphi_1$. The associated beta function only 
depends on $z_k$, $u'_k$, $\delta_{k,0}$, $\delta_{k,2}$ and their (field) derivatives. In this case, as already mentioned, 
$\delta_{k,0}(\varphi)=z_k(\varphi)$ and the flows of $z_k$ and $u'_k$ are the same as in the pure system in dimension 
$d-2$ (with $d=3$ here) at the same level of the derivative expansion.

We have therefore solved the two coupled partial differential equations for $z_k$ and $u'_k$ numerically in $d=1$, 
for a range of $\sigma$ between $1/2$ and $1$, and we have taken the result as an input to solve the 
partial differential equation for $\delta_{k,2}(\varphi)$. For the reduced cutoff function $s(y)$, we have used, 
as in our previous work,\cite{tissier06,tissier12b} $s(y)= a(1+y/2 +y^2/12)\exp(-y)$, 
where the parameter $a$ has been optimized via stability considerations\cite{litim00,canet03,pawlowski07} 
and has been varied to provide an estimated error bar on our results (in practice, stable results are 
obtained for a wide range of $a$, between 1.5 and 6).

\begin{figure}[ht]
\includegraphics[width=\linewidth]{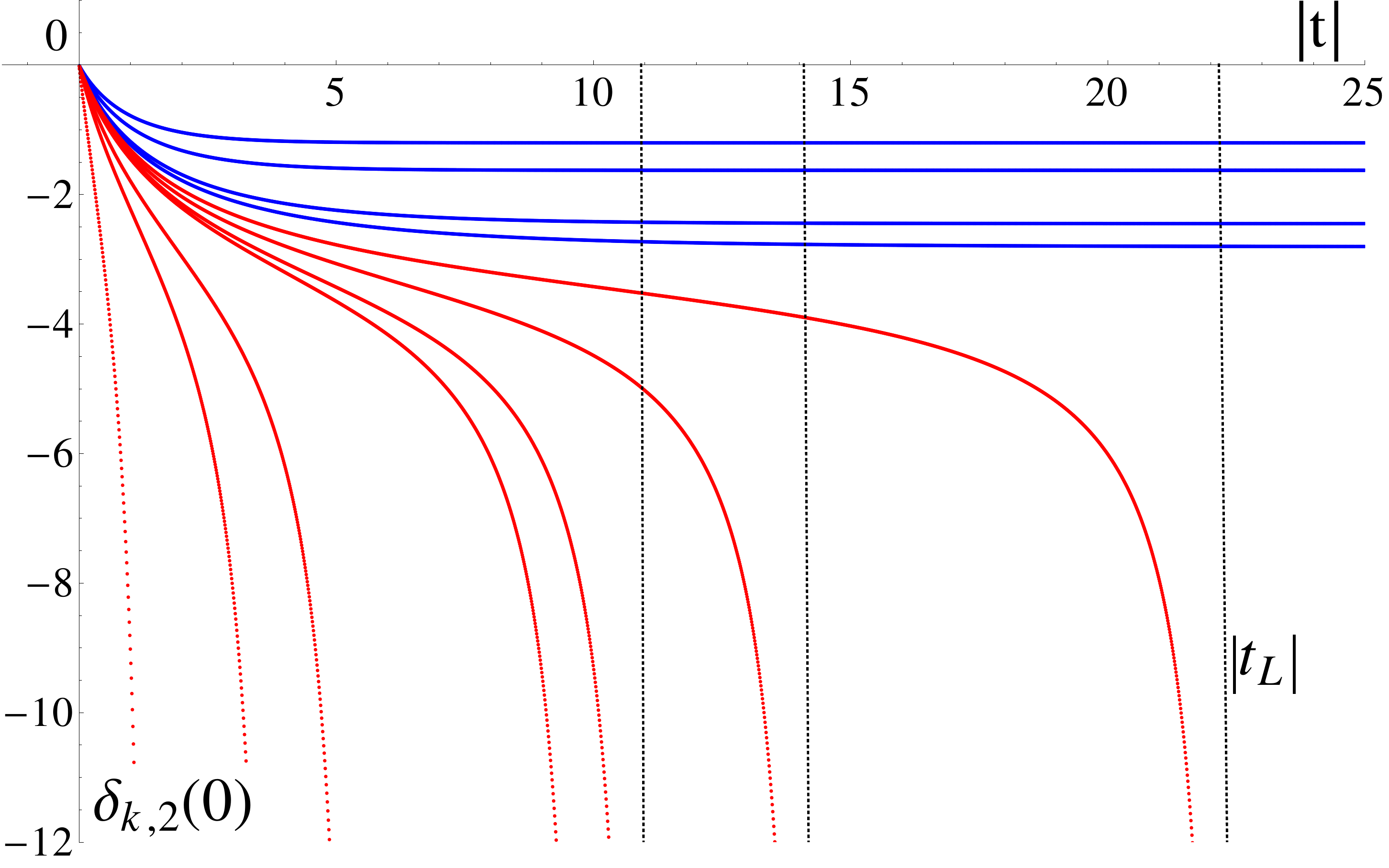}
\caption{\label{fig_delta2} NP-FRG flow of $\delta_{k,2}(\varphi=0)$ in the $3$-dimensional long-range RFIM. The initial conditions 
at $k=\Lambda$ (\textit{i.e.}, $t=0$) for $u'_k(\varphi)$ and $z_k(\varphi)=\delta_{k,0}(\varphi)$ are taken as equal 
to the solution at the fixed-point. The upper (color online blue) curves correspond to $\sigma<\sigma_{DR}\approx 0.72$: 
$\sigma = 0.7, 0.71, 0.72, 0.722$ 
from top to bottom; $\delta_{k,2}(0)$ tends to a finite fixed-point value. The lower (color online red) curves 
correspond to $\sigma>\sigma_{DR}$: $\sigma=0.8, 0.75, 0.74, 0.73, 0.729, 0.727, 0.725$ from left to right; they all display a 
divergence at a finite RG scale $\vert t_L\vert$ (which is indicated for the three rightmost curves by a vertical dashed line). 
Here, the value of the parameter $a$ of the cutoff function is equal to 2.}
\end{figure}

\begin{figure}[ht]
\includegraphics[width=\linewidth]{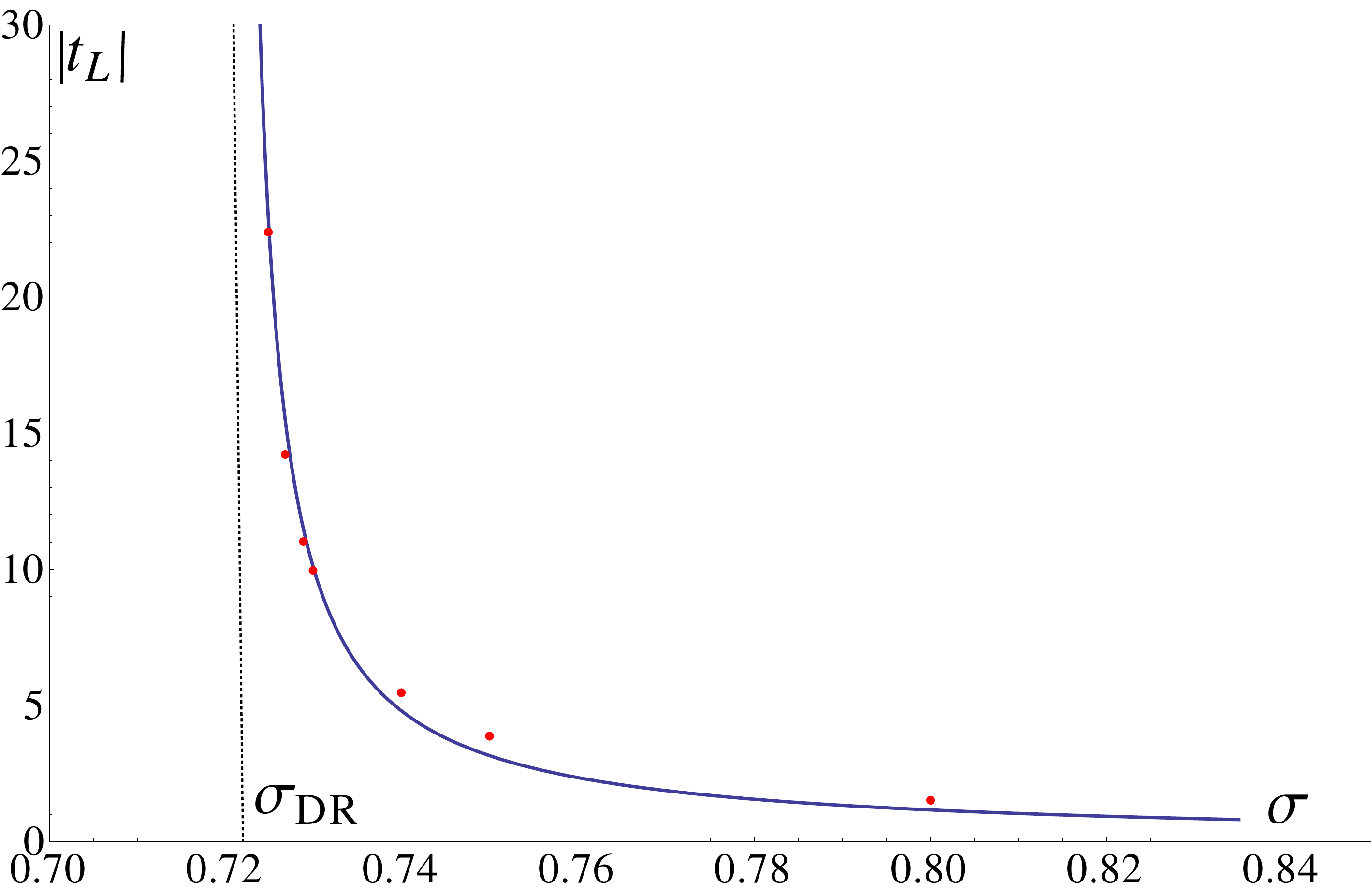}
\caption{\label{fig_larkin} Divergence of the ``Larkin'' RG time $\vert t_L\vert$ when $\sigma\rightarrow \sigma_{DR}^+$. 
The line is a simple power law in $(\sigma_{DR}-\sigma)^{-1}$. (note that an RG time of 23 is enormous and corresponds 
to an observation length scale $10^{10}$ times the microscopic one).}
\end{figure}

The outcome of our theoretical investigation is that dimensional reduction between the RFIM with long-range interactions 
and disorder correlations in $d=3$ and the pure model with long-range interactions in $d=1$ is valid for $\sigma<\sigma_{DR}$ 
and breaks down for $\sigma>\sigma_{DR}$, with a critical value $\sigma_{DR} \approx 0.72 \pm 0.03$. Indeed, and as 
illustrated in Fig.~\ref{fig_delta2}, $\delta_{k,2}(\varphi)$ stays finite down to $k\rightarrow 0$ 
below $\sigma_{DR}$ whereas it diverges at a finite scale above $\sigma_{DR}$. We also display in Fig.~\ref{fig_larkin} the divergence of 
the ``Larkin'' scale at which $\delta_{k,2}(\varphi=0)\rightarrow \infty$ when 
$\sigma\rightarrow \sigma_{DR}^+$.

We stress that dimensional reduction breaks down above $\sigma_{DR}$ despite the fact that 
$\bar \eta=\eta=2-\sigma$ (as in the pure model) and $\theta=2$. In this case indeed, the correlation length 
exponent $\nu$ differs from its value in the pure 
long-range system in two dimensions less. We have not explicitly computed $\nu$ at this stage, as it 
requires the resolution of the full set of coupled differential equations, including a function $\delta_k(\varphi_1,\varphi_2)$ 
depending on two fields and displaying a cusp in $\vert \varphi_1-\varphi_2\vert$, which represents a 
much harder numerical task that we postpone to future investigation. However, one can show that the dimensional reduction fixed point 
disappears in the presence of a cusp and that the exponent $\nu$ in the presence of a cusp has additional contributions not present in 
the dimensional reduction case: $\nu_{RFIM}(d)\neq \nu_{Ising}(d-2)$. On the other hand, below $\sigma_{DR}$ 
we rigorously find $\nu_{RFIM}(d)=\nu_{Ising}(d-2)$.

According to our recent work, the validity of dimensional reduction in disordered systems controlled by a zero-temperature 
fixed point is related to the scaling properties of the large avalanches observed in the evolution of the ground state under 
changes of the external source.\cite{tarjus13} Avalanches are the physical source of the appearance of a 
nonanalyticity in the cumulants of the renormalized disorder. Dimensional reduction is broken when the avalanche 
contributions are relevant at the fixed point 
and is valid when avalanches lead to subdominant effects. In the former case ($\sigma>\sigma_{DR}$), the scaling 
dimension $d_f$ of the largest typical system-spanning avalanches at criticality is equal to $(d+4-\bar \eta)/2$, 
\textit{i.e.} to $(5+\sigma)/2$ in the present study; on the other hand,  for $\sigma<\sigma_{DR}$,  $d_f=(5+\sigma)/2-\lambda$,
where $\lambda >0$ is the eigenvalue associated with a ``cuspy'' perturbation to the cuspless dimensional-reduction 
fixed point.\cite{tarjus13} The eigenvalue $\lambda$ increases as $\sigma$ decreases. It is easily derived that 
$\lambda=(5-\sigma)/2$ around the gaussian fixed point (for $\sigma\leq 1/2$), so that $d_f=1$ for $\sigma=1/2$ [while 
$(5+\sigma)/2=2.75$].

Finally, we conclude this presentation by stressing the relevance to computer studies, either simulations or exact ground-state 
determinations, which represents the main motivation of this work. In such studies, one considers a lattice, ``hard-spin'' version 
of the RFIM, which has the following Hamiltonian in the long-range case:
\begin{equation}
\label{eq_LR_RFIM}
H=-\frac 12 \sum_{i,j=1}^N J_{ij}S_iS_j -\sum_{i=1}^N h_i S_i
\end{equation}
where $S_i=\pm1$, the interaction term $J_{ij}\equiv J(\vert x_i-x_j\vert)$ goes as $\vert x_i-x_j\vert^{-(d+\sigma)}$ 
at large distance, and the random fields $h_i$ are Gaussian distributed with zero mean and long-range correlator 
$\overline{h_i h_j }=\Delta(\vert x_i-x_j\vert) \sim \vert x_i-x_j \vert^{-(d-2+\sigma)}$. To ensure that the corresponding generating functional 
satisfies superrotational invariance, the following relation, which is a variant of the Ward-Takahashi equation in 
Eq.~(\ref{eq_ward-takahashi}),  should hold:
\begin{equation}
\label{eq_WardLR_RFIM}
\frac{d}{dr}\Delta(r)=- C\, r J(r) \, ,
\end{equation}
where $C$ parametrizes the relative strength of the disorder compared to the interactions. Then, powerful algorithms exist 
to determine the exact ground state of a sample in a computer time that scales 
only as a power of the system size.\cite{dauriac85,ogielski86} This allows one 
to investigate large systems and to perform efficient finite-size analyses of the critical behavior at 
zero temperature.\cite{middleton,hartmann,frontera00,machta,liu07} The algorithms can be extended to describe long-range correlated 
disorder\cite{hartmann11} and long-range interactions,\cite{frontera01} and for the cubic lattice, large system sizes 
should be attainable. A systematic study of the model as a function of the parameter $\sigma$ could therefore provide the 
first direct independent check of our theoretical predictions concerning dimensional reduction and its breakdown.

\end{document}